\documentclass[aps, pre, reprint]{revtex4-1}
\usepackage{graphicx}
\usepackage{float}
\usepackage{graphics}
\usepackage{color}

\usepackage{amsmath}
\usepackage{amsfonts}

\begin{document}
%
\title{Thermodynamic properties and Hilbert space of the human brain}

\author{Dongmei Shi$^{1,2,3}$}
\author{Meng Li$^{4}$}
\author{Martin Walter$^{4}$}
\author{Hamid R.~Noori$^{1,5,6}$}
\affiliation{
$^1$ Max Planck Institute for Biological Cybernetics, T\"ubingen 72076 Germany,\\
$^2$ Institute for Neuromodulation and Neurotechnology, T\"ubingen University Hospital , T\"ubingen 72076, Germany,\\
$^3$ Department of Physical Science and Technology, Bohai University, Jinzhou 121000, China,\\
$^4$ Department of Psychiatry and Psychotherapy, Jena University Hospital, Jena 07743, Germany\\
$^5$ McGovern Institute for Brain Research, Massachusetts Institute of Technology
Cambridge, MA 02139, USA,\\
$^6$ Laboratory of Multiscale Dynamics of Brain Disorders, Center for Excellence in Brain Science and Intelligence Technology Institute of Neuroscience, Chinese Academy of Sciences, Shanghai 200031, China\\
}

\begin{abstract}
Any macrosystem consists of many microparticles. According to statistical physics, the macroproperties of a system are realized as the statistical average of the corresponding microproperties. In our study, a model based on ensemble theory from statistical physics is proposed. Specifically, the functional connectivity (FC) patterns confirmed by Leading Eigenvector Dynamics Analysis (LEiDA) are taken as the microstates of a system, and static functional connectivity (SFC) is seen as the macrostate. When SFC can be written as the linear combination of these FC patterns, it is realized that these FC patterns are valid microstates for which the statistical results of relevant behaviors can describe the corresponding properties of SFC. In this case, the thermodynamic functions in ensemble theory are expressed in terms of these microstates. We apply the model to study the biological effect of ketamine on the brain and prove by {\it maximum work principle} that compared to that in the control group, the capability of work done by the brain that has been injected with ketamine declines significantly. Moreover, the quantum mechanical operator of the brain is further studied, and a Hilbert space spanned by the eigenvectors of the operator is obtained. The confirmation of the mechanical operator and Hilbert space opens great possibilities of using quantum mechanics to study brain systems, which would herald a new era in relevant neuroscience studies.

\end{abstract}

\maketitle


%

\section{Introduction}
%
%
%
%
The human brain is a complex system consisting of functionally and structurally interconnected regions. This organ is an advanced functional system that performs diverse cognitive functions in a constantly changing environment \cite{brain1,brain2}. Even when no task is performed, the brain displays the spontaneous waxing and waning of meaningful functional networks on a slow time scale \cite{rest1,rest2,rest3}. Understanding the relation between brain architecture and function is a central challenge in neuroscience.

In recent decades, structural and functional neuroimaging studies have provided much knowledge about primate and human brains~\cite{rest2,neuim1,neuim2,neuim3,neuim4,neuim5,neuim6}. In addition, computational network analysis has provided insight into the organization of large-scale cortical connectivity in several species, including rats, cats and macaque monkeys~\cite{cnet1,cnet2,cnet3,cnet4}. 
In the human brain, the topology of functional connectivity (FC) patterns has been widely investigated~\cite{FC1,FC2,FC3,FC4},  and key attributes of these patterns have been characterized across different conditions of rest or cognitive load. FC~\cite{FCdef1,FCdef2} is defined as the temporal dependence of neuronal activity patterns of anatomically separated brains. Many studies have shown the feasibility of examining FC between brain regions as the level of coactivation of functional MRI time series measured during rest~\cite{FCdef3,FCdef4}. Resting-state activity has been proposed to reflect the spontaneous activation and deactivation of different network configurations~\cite{rest4,rest5,rest6,rest7}, resulting in a constant reconfiguration of FC patterns over time. However, this pattern cannot be captured by traditional static functional connectivity SFC analysis, for which BOLD signal correlation is computed over the entire recording session~\cite{asfc1,asfc2}. New methods have been proposed to define FC patterns \cite{FCdef4,newm1,newm2,newm3,newm4,newm5}, and among them phase coherence connectivity enables a higher temporal resolution, but is more susceptible to high-frequency noise fluctuations. To overcome this shortcoming of phase coherence connectivity, J. Cabral $et$ $al$~\cite{cabral} focused on the dominant FC patterns captured by the leading eigenvector of BOLD phase coherence matrices, investigating how the switching behavior of resting-state FC patterns is related to cognitive performance. By adopting the same approach LEiDA to define FC, Deco $et$ $al$ \cite{neuim6} proposed a possible definition of brain state, i.e., an ensemble of 'metastable substates', to study the transition of brain states.

Inspired by the work of Cabral \cite{cabral} and Deco \cite{neuim6}, we proposed a model in which the FC patterns are defined as the microstates and SFC is seen as the macrostate of a system. When the linear relation between SFC and FC patterns is satisfied, the thermodynamic properties of SFC can be described by FC patterns via ensemble theory. In Deco's \cite{neuim6} work, a brain state is defined as an ensemble of FC patterns (or 'metastable substates') from a macroscopic scale, whereas in the present study, it is defined as an FC pattern from the microscopic scale. These two definitions can be connected by statistical physics, which provides disciplines to describe the macroproperties of a system by analyzing the corresponding microproperties, which is the basic idea of our model. We applied the model to study the effect of ketamine on the human brain and compared the results with those of the control group. Significant differences were found between the two groups. In addition, work done and heat conduction occurring in the brain are studied by means of the {\it first law of thermodynamics}. We further investigated the mechanical operator of the brain system and built a Hilbert space spanned by the eigenvectors of the operator. This work provides new research perspectives for studying neuroscience, from which thermodynamics, statistical physics and quantum mechanics could be applied. 

The data adopted in our work were collected from 31 participants from a ketamine group and 30 participants from a control group in which placebo was injected. Each participant experienced three stages: baseline, at which the subjects had not been injected with ketamine (or placebo); stage 1, at which the subjects had been injected with ketamine (or placebo) after one hour; and stage 2, at which the subjects had been injected with ketamine (placebo) after 24 hours.

\section{Experimental Data}

\begin{enumerate}
\item {\textsf {Study design and Participants}}

The data set was collected from a completed randomized, double-blind, placebo-controlled study at the University of Magdeburg (Magdeburg, Germany), which was designed to investigate the biological effects of ketamine (more details in the registered clinical trial at the European Union Drug Regulating Authorities Clinical Trials (No. 2010-023414-31)).
The study design has been reported in our previous studies \cite{data1,data3}. In brief, three measurements, namely, at baseline and at 1 h and 24 h post-infusion, were collected for each participant. Following the baseline fMRI scan, each participant was randomized by age and sex to a 40 min continuous infusion of either (R,S)-ketamine (0.5 mg$/$kg) or 0.9$\%$ saline via an infusion pump outside the MRI scanner. The second MR scan was started at 1 h post-infusion, and the resting-state fMRI scan was performed approximately 100 min post-infusion. The third fMRI scan was acquired 24 h after their associated baseline measurements. During ketamine administration, participants may experience transient dissociation and hallucination, as well as increased heartrate and blood pressure. The half-life of ketamine is approximately 80 min. Here, the proposed framework was applied to investigate the dynamics of brain states and the mechanical operator of the evolution from baseline to reconfiguration and normalization at the end of this study.

A group of 80 healthy volunteers (33 female, mean age $\pm$ SD: 25.9 $\pm$ 5.3 years) was recruited and screened by mini-international neuropsychiatric interviews (MINI, German Version 5.0.0), physical and ophthalmic examinations, blood laboratory tests, and electrocardiography in Magdeburg, Germany. All participants were in good physical health and had no pregnancy, left-handedness, or any MRI-incompatible devices. A current or former history of substance use disorder was excluded, and drug screening was performed.
The study was approved by the Institutional Review Board of the Otto-von-Guericke-University Magdeburg, Germany. All subjects gave written informed consent and received financial reimbursement.

\item {\textsf {MRI scan and resting-state functional connectivity calculation } }

All scans were acquired on a 7 T MR scanner (Siemens Healthcare, Erlangen, Germany) using a 32-channel head array coil (for RS-fMRI: 280 time points, TE $=$ 22 ms, TR $=$ 2.8 s, flip angle $= 80^\circ$, 62 contiguous axial slices, bandwidth = 2246 Hz/pixel, acquisition matrix = 106 $\times$ 106, voxel size $= 2 \times 2 \times 2~mm^{3}$; T1- weighted MRI: 3 D-MPRAGE sequence, TE $=$ 2.73 ms, TR = 2300 ms, T1 = 1050 ms, flip angle = $5^\circ$, bandwidth = 150 Hz$/$pixel, acquisition matrix = $320 \times 320$, voxel size = $0.8 \times 0.8 \times 0.8~mm^{3}$).

To exclude anatomical abnormalities and abnormal frontal signal dropout, individual EPI data were visually inspected by two independent raters, and an additional point-spread function (PSF) mapping protocol was scanned for EPI distortion correction. Twenty-three EPI scans out of 240 total scans were discarded and resulted in 19 subjects (7 female, age 27.1 $\pm$ 6.9 years) with a low-quality EPI in one or more sessions.

Next, the pipeline developed in the 1000 Functional Connectomes Project (https://www. nitrc.org/projects/fcon$\_$1000/) with few modifications was performed for data preprocessing. The major steps of this well-established pipeline include brain extraction, time slicing, realignment, nuisance regression, normalization, temporal filtering with a bandpass filter (0.01-0.1 Hz), and spatial smoothing (FWHM 6 mm). The framewise displacement was calculated and added as an additional confounding factor along with the motion parameters (three translations, three rotations), as well as the mean signals of white matter and cerebrospinal fluid (CSF), using a multiple linear regression for each subject. With the proposed image in standard MNI space, the parcellation scheme with 200 ROIs by Schaefer and colleagues \cite{data2} was applied in this study. In each subject, the average time course within each parcel was extracted, and a 200$\times$200 adjacent matrix was created.

\end{enumerate}

\section{Methods}

\subsection{Confirming the brain state by LEiDA}

The functional connectivity FC patterns are defined as the microstates of the brain and determined by the LEiDA \cite{cabral,neuim6}.
Specifically, the dynamic functional connectivity (dFC) is first calculated, which is a phase coherence matrix and defined as
\begin{equation}
dFC (n,p,t) = cos (\theta (n,t) - \theta(p,t)).
\end{equation}

$n$ (or $p$) denotes the brain area $n$ (or $p$), and $\theta (n, t)$ (or $\theta (p, t)$) represents the phase function of brain area $n$ (or $p$) at time point t. At each point in time, each brain area has a corresponding phase function that is obtained by the Hilbert transform~\cite{newm1,newm2,newm3,newm4,newm5}. Apparently, the dFC is a dynamic function describing the functional coherence between the brain areas at each time point.

Considering that the dFC at each time point is undirected and symmetric across the diagonal, the dimension of dFC can be largely reduced by calculating its leading eigenvector $v$, which captures the instantaneous dominant connectivity pattern of dFC. Upon computing the leading eigenvector of the dFC, the recurrent FC patterns are identified by applying the $k-$means clustering algorithm~\cite{km1,km2,km3,km4}. For further details regarding LEiDA, please read \cite{cabral}.

\subsection{Equilibrium ensemble theory model}

Compared to dynamic functional connectivity dFC, static functional connectivity (SFC), for which the correlation between the brain areas is computed over the entire recording session, describes an average effect of the functional coherence between brain areas. The FC patterns describe the functional coherence between the brain regions during the recording session, whereas SFC describes the average effect of the functional correlation between the brain areas, which permits us to take SFC as the statistical average of the relevant performance of FC patterns.

{\bf Assumptions:} In our study, it is realized that the FC patterns obtained by LEiDA are valid to present the statistical properties of SFC when the similarity shown in equation (2) is no less than $80\% $. 
$w_{i} = v_{i}*v_{i}^{T}$ is the outer product of FC pattern $v_{i}$, and $p_{i}$ is the corresponding occurrence frequency. $S_{imi}$ denotes the similarity that is measured as the Pearson correlation ($P_{corr}$) between SFC and the weighted sum of $w_{i}$. Before we introduce the ensemble theory model, some assumptions must be presented.

\begin{equation}
	S_{imi}= P_{corr} (SFC, \sum_{i} p_{i}w_{i})
\end{equation}

\begin{enumerate}

\item The resting state of the brain is taken as an equilibrium at which the brain does not process any externally promoted tasks. Macroscopically, the brain state is static (e.g., sleeping); however, microscopically, brain activities are always present.

\item FC patterns are defined as the microstates of the brain, whereas SFC is taken as the brain's macrostate.

\item When the condition $S_{imi} \geq 0.80 $ meets, $ \{v_{i}, p_{i}  | i = 1,...,n\}$ corresponds to the valid microstates of which the statistical average of performance can describe the properties of macrostate SFC.

\item The leading eigenvalue of $w_{i} = v_{i}*v_{i}^{T}$ is set to represent the system's $E_{i}$ when the system stays in state $v_{i}$.

\end{enumerate}

For the assumption of energy $E$, we have tried three different parameters. In addition to the leading eigenvalue of $v*v^{T}$, other parameters, including the average eigenvalue of dFCs' leading eigenvectors that belong to the same cluster and the average of corresponding Bold signals, are evaluated. It has been proven that these three parameters generally show consistent results. Without losing generality, the leading eigenvalue of $v*v^{T}$ is finally taken as the system's energy in our model.

{\bf Equilibrium ensemble theory (EET)}: In statistical physics, an ensemble is an idealization consisting of a large number of virtual copies of a real system. Considering all the copies at once, each copy represents a possible microstate that the real system might be in. A thermodynamic ensemble studies a system when it remains in statistical equilibrium~\cite{ensem1,ensem2,ensem3}. When the system is realized to have determined particles N, volume V and temperature T, {\it canonical ensemble theory} is applicable, and the corresponding partition function $Z$ is written as

\begin{equation}
	Z= \sum_{i=1} e^{-\beta E_{i}}.
\end{equation}

where $E_{i}$ is the energy of the system when the system remains in the microstate $v_{i}$. $\beta = \frac{1}{\kappa T}$, where $\kappa$ is the Boltzmann constant $1.38 * 10^{-23}~J/K$ \cite{boltz_c} and T is the absolute temperature with the unit {\it K} \cite{kelwin}.

Considering that $\kappa$ is very small and $E_{i}$ is also a small value (which is represented by the leading eigenvalue and normally $< 3$), to make $lnZ$ meaningful, we have modified $\kappa$ to another constant 
$R = \kappa N_{a} \rho = 8.31~J/K$. $N_{a}$ is the Avogadro constant ($6.02*10^{23} / mol$), and $\rho = 1$ mol. The constants $\kappa$ and $R$ have consistent units. Accordingly, the partition function is rewritten as

\begin{equation}
	Z= \sum_{i=1} e^{-\gamma E_{i}},
\end{equation}
where $\gamma = \frac{1}{RT}$. The partition function $Z$ describes the statistical properties of the system, based on which other thermodynamic functions can be presented. Accordingly, the internal energy $U$, thermodynamic entropy $S$ and Helmholtz free energy $F$ are written as

\begin{equation}
	U= - \frac{\partial lnZ}{\partial \gamma},
\end{equation}

\begin{equation}
	S= R(lnZ - \gamma \frac{\partial lnZ}{\partial \gamma}),
\end{equation}

and

\begin{equation}
	F= - RTlnZ.
\end{equation}

\section{Results and Discussion}

\subsection{Thermodynamic properties of the brain}

Based on the data provided (see 'Experimental Data'), we use phase coherence connectivity to obtain the dFC, with the size $N \times N \times T$, where $N= 200$ is the number of parcellated brain regions, and $T= 276$ is the number of recording frames. The microstates are confirmed by performing $k-means$ clustering analysis on all the leading eigenvectors across time points and subjects (i.e., $276\times 31 = 8556$ leading eigenvectors for the ketamine group). By doing so, as many significant microstates as possible are collected, and the statistics are therefore more meaningful. 
The parameter $k$ in $k-means$ clustering changes from 2 to 50, repeating each 30 times. Each thermodynamic property obtained below is the average over these 30 calculations.

$S_{imi}$ is first calculated for both groups. It is found that $\overline{S_{imi}} \ge 0.80 $ in each stage, which means that $SFC$ can be presented by the linear combination of the microstates. Furthermore, the brain system at each stage can be seen with determined particles $N$, volume $V$ and temperature $T$; accordingly, {\it canonical ensemble theory} based on the EET model described above is applied. $T$ is further seen not to change across the stages and is set to 1~$K$ without losing generality.

{\bf The thermodynamic properties} The internal energy $E$, entropy $S$ and $F$ for the ketamine group and the control group are shown in Table 1 and Table 2, respectively.

In Table 1, one can see that compared to that at baseline, the internal energy does not change significantly, whereas the entropy has increased in stage 1 and stage 2, and the free energy has decreased. According to the Boltzmann formula of entropy, the entropy can also be written as $S = \kappa ln\Omega$, where $\Omega$ corresponds to the number of microstates. Therefore, one can conclude that more microstates in the brain emerged after the injection of ketamine. Table 2 shows the corresponding results for the control group. One can see that the internal energy, entropy and free energy show similar trends with those observed in the ketamine group. 

Apparently, although the entropy or the free energy changed after the subject was injected with ketamine, we cannot say that these changes were caused by ketamine since similar changes also occurred in the control group. Therefore, the work done by the brain is further investigated.

{\bf Maximum work principle} According to the {\it maximum work principle} \cite{mwp1}, when a system experiences an isothermal process from state 1 to state 2, the maximal work done by this system is the decrease in free energy, i.e.,

\begin{equation}
\nabla = F_{1} - F_{2} \geq -W.
\end{equation}

$-W$ is the work done by the system.
For our case, during the process from baseline to stage 1 or stage 2, the temperature of the brain or the environment does not exhibit very significant changes; thus, the process can be seen as an isothermal process. According to the results shown in the above tables, we obtain that for the ketamine group, the decrease in free energy between baseline and stage 1 is $\nabla_{ketamine}^{1} = F_{base} - F_{s1} = 1.3 $, and that between baseline and stage 2 is $\nabla_{ketamine}^{2} = F_{base} - F_{s2} = 1.71$; for the controlled group, the corresponding results are $\nabla_{placebo}^{1} = 4.63$ and $\nabla_{placebo}^{2} = 3.41 $. Thus, it can be concluded that the maximal work that the brain can do as measured in the ketamine group is significantly less than that as measured in the controlled group. If we understand the maximal work that the brain can do as the work capability of the brain to perform tasks, then the results tell us that the work capability of the brain measured in the ketamine group is obviously weaker than that in the control group.

\begin{table*}[!ht]
	\caption{Thermodynamic Functions of Ketamine group}
	\newcommand{\tabincell}[2]{\begin{tabular}{@{}#1@{}}#2\end{tabular}}
	\centering
	\begin{center}
		\begin{tabular}{||c | c | c | c||}
			\hline
			{Stages} : {Functions} & Internal Energy $ \overline{U}$ & Entropy $\overline{S}$ & Free energy $\overline{F}$ \\ [0.5ex]
			\hline\hline
			Baseline & 0.79 & 27.33 & -26.54 \\
			\hline
			Stage 1  &0.78 & 28.63 & -27.84\\
			\hline
			Stage 2 & 0.79 & 29.04 & -28.25 \\
			[1ex]
			\hline
		\end{tabular}
	\end{center}
\end{table*}

\begin{table*}[!ht]
	\caption{Thermodynamic Functions of Placebo group}
	\newcommand{\tabincell}[2]{\begin{tabular}{@{}#1@{}}#2\end{tabular}}
	\centering
	\begin{center}
		\begin{tabular}{||c | c | c | c||}
			\hline
			{Stages} : {Functions} & Internal Energy $ \overline{U}$ & Entropy $\overline{S}$& Free energy $\overline{F}$ \\ [0.5ex]
			\hline\hline
			Baseline & 0.79 & 25.90 & -25.14 \\
			\hline
			Stage 1  & 0.80 & 30.57 & -29.77\\
			\hline
			Stage 2 & 0.78 & 29.33 & -28.55 \\
			[1ex]
			\hline
		\end{tabular}
	\end{center}
\end{table*}

\subsection{Heat conduction of the brain}

In this part, we studied the heat conduction occurring in the brain in the ketamine group.

From the first law of thermodynamics,

\begin{equation}
	\triangle{U}= Q + W.
\end{equation}

where $U$ is the internal energy, and $\triangle {U}$ denotes the change in $U$. $Q$ denotes the heat that the system has absorbed, and $W$ indicates the work that the environment has done to the system. From the result in part A, the internal energy does not change significantly, so we have
$\triangle{U}=0$. Accordingly,

\begin{equation}
	W = -Q.
\end{equation}

From the Clausius entropy expression \cite{clau1,clau2,clau3,clau4}, for any reversible process from state {\it a} to state {\it b}, the change in entropy $\triangle{S}$ is expressed as

\begin{equation}
	\triangle{S} = S_{b}-S_{a} = \int_{a}^{b}\frac{dQ}{T}. 
\end{equation}

Since the entropy S is a state function, the change $\triangle{S}$ depends only on the initial state and terminal state. Therefore, theoretically, we can always create a reversible process that has the same initial state and same terminal state and then apply the formula to study $\triangle{S}$. Let us take baseline as the initial state $a$, ketamine stage (stage 1 or stage 2) as the terminal state $b$, and $T$ as invariable during the reversible process. Then, the change $\triangle{S}$ is

\begin{equation}\nonumber
	\triangle{S} = \int_{a}^{b}\frac{dQ}{T} = \frac{1}{T}\int_{a}^{b}dQ = \frac{Q}{T},
\end{equation}
and we thus have

\begin{equation}\nonumber
	Q = T \triangle{S}.
\end{equation}

From the results of S in stage 1 or stage 2, we have $\triangle {S} >0$. Together with equation $(10)$,  we obtain
\begin{equation}\nonumber
	Q >0,  W (= -Q) <0 .
\end{equation}
The results indicate that during the process from baseline to the ketamine stage, the brain absorbs heat from the external environment and does positive work to the external environment. The environment includes any other systems that interact with the brain system.

\subsection{Discussions}

From Table 1, one can see that although some thermodynamic properties have changed, the changes are not significant. For example, the entropy S increases by 1.3 in stage 1 and 1.71 in stage 2 compared to that at baseline. Since SFC represents a macrostate, we further studied the similarity between the SFCs from any two different stages. The similarity is measured as the Pearson correlation. It is found that the similarity between SFC from baseline and SFC from any other stage is $\approx 0.95$, which means that the macrostates of these three stages are all like, so the macroproperties from different stages do not have distinct differences. The reason why it shows a similar macrostate at each stage may be that the ketamine concentration in the body has largely decreased even after one hour, so the brain has recovered to a large extent. However, through the {\it maximum work principle}, a significant difference between the ketamine group and the control group was presented by the decrease in $F$, despite the large similarity between macrostates.

\section{Quantum mechanical operator and Hilbert space of the brain}
\subsection{Confirming the mechanical operator $\mathcal{A}$}
In quantum mechanics, a mixed quantum state $\Phi$ can be described as
\begin{equation}
	\Phi \to
	\begin{cases}
	\phi_{1}, & f_{1} \\
	\vdots & \vdots \\
	\phi_{i}, & f_{i} \\
          \vdots & \vdots \\
	\phi_{k},& f_{n}\\
	\end{cases}
\end{equation}
where $\phi_{i}$ is the ${\it i}$th pure state, and $f_{i}$ is the corresponding probability with which the system stays in $\phi_{i}$. $f_{i}$ meets

\begin{equation}
	\sum_{i}f_{i}=1
\end{equation}

In our study, the FC states $\{v_{i}|  i=1,...,n\}$ are taken as the microstates, and the corresponding probabilities $\{p_{i}|  i=1,...,n\}$ meet the condition shown in equation (13). Let us assume that there exists a mixed quantum state that is described by the pure state $\{v_{i}|  i=1,...,n\}$ and that there is a mechanical operator $\mathcal{A}$ correlated to $\{v_{i}|  i=1,...,n\}$. To confirm this operator, we have the following hypotheses:

\begin{enumerate}
\item 	The pure states $\{v_{i}|  i=1,...,n\}$ have close relations with $\mathcal{A}$'s characteristic functions $\Psi = \{ \psi_{i} | i=1,...,n\}$ . When n = 200 (the number of parcellated brain regions), the characteristic functions of A can be obtained by the Schmidt process performed on $\{v_{i}|  i=1,...,n \}$.

\item  	The system's energies  $\{E_{i}|  i=1,...,n\}$ when the system stays in $\{v_{i}|  i=1,...,n \}$ have close relations with the characteristic values $\Lambda$ of $\mathcal{A}$ . When n = 200, $\{E_{i}|  i=1,...,n\}$ are the corresponding characteristic values $\Lambda =\{\lambda _{i} = E_{i}  | i=1,...,200\}$.

\end{enumerate}

Based on the above analyses, we confirm the mechanical operator $\mathcal{A}$ by the following steps.
\begin{enumerate}

\item{\textsf{Step 1}}: We look for a mixed state $\Phi$ that has 200 pure states. Moreover, the relation of $\overline{S_{imi}}$ no less than 0.8 still holds.

\item {\textsf{Step 2}}: We verify whether these 200 states $\{v_{i}|  i=1,...,200\}$ are linearly uncorrelated.
If they are linearly uncorrelated, we confirm the characteristic functions $\Psi$ by performing the Schmidt process on the states $\{v_{i}|  i=1,...,200\} $ to obtain the characteristic functions $\{\psi_{i} | i=1,...,200\}$; otherwise, if they are linearly correlated, we return to step 1 until all 200 pure states are linearly uncorrelated.

\item {\textsf{Step 3}}: According to linear algebra, the mechanical operator $\mathcal{A}$ can be determined by
\begin{equation}
	\mathcal{A}  = R_{A}U_{\Lambda}R_{A}^{-1},
\end{equation}
where $R_{A}$ is a matrix (full rank) whose columns are the different vectors $\{\psi_{i} | i=1,...,200\}$ from $\Psi$; $U_{\Lambda}$ is a diagonal matrix in which the diagonal elements are the corresponding eigenvalues; and $R_{A}^{-1}$ is the inverse matrix of $R_{A}$.

\end{enumerate}

By applying equilibrium ensemble theory, we have
\begin{enumerate}

\item In each stage, a mixed state $\Phi$ with 200 linearly uncorrelated pure states is found, and accordingly, the corresponding mechanical operator is confirmed for each stage.
\item Let us write the mixed states of baseline, stage 1 and stage 2 as $\Phi_{0}$, $\Phi_{1}$ and $\Phi_{2}$, respectively, and write the corresponding mechanic operators as $\mathcal{A}_{0}$, $\mathcal{A}_{1}$ and $\mathcal{A}_{2}$. It has been confirmed that for any ${\it i,}$ ${\it  j}$, and $i \neq j (i, j \in 0,1,2)$,
\begin{equation}\nonumber
\Phi_{i} \neq \Phi_{j}, \mathcal{A}_{i} \neq \mathcal{A}_{j}.
\end{equation}
\end{enumerate}

In short, the operators are confirmed for each stage, and any two of the operators are different. In the following, we will discuss the relations of the three mechanical operators.

\subsection{The relation of the mechanical operators between different stages}
\subsubsection{Similarity matrix of the mechanic operator}

First, let us see the operator $\mathcal{A}_{0}$ when the brain remains at baseline.
$\mathcal{A}_{0}$ is obtained by
\begin{equation}\nonumber
\mathcal{A}_{0}= R_{0}U_{0}R_{0}^{-1},
\end{equation}
so from $\mathcal{A}_{0}$, we have

\begin{equation}
U_{0}= R_{0}^{-1}\mathcal{A}_{0}R_{0}.
\end{equation}

According to linear algebra, two $n\times n$ matrices M and N are considered similar if there exists an invertible $n \times n$ matrix $\Gamma$ such that
\begin{equation}
N= \Gamma^{-1} M \Gamma .
\end{equation}

Therefore, based on equation (15) of $U_{0}$, it is concluded that $U_{0}$ is the similarity matrix of $\mathcal{A}_{0}$, i.e.,
\begin{equation}\nonumber
\mathcal{A}_{0} \sim U_{0}.
\end{equation}

Similarly, in stage 1 and stage 2, we have
\begin{equation}\nonumber
\mathcal{A}_{1} \sim U_{1}, \mathcal{A}_{2} \sim U_{2},
\end{equation}
where $U_{1}$ and $U_{2}$ are the corresponding diagonal matrices of $\mathcal{A}_{1}$ and $\mathcal{A}_{2}$, respectively.

\subsubsection{The relation of eigenvalue spectra of $\mathcal{A}_{0}$, $\mathcal{A}_{1}$ and $\mathcal{A}_{2}$}

The eigenvalue spectra of $\mathcal{A}_{0}$, $\mathcal{A}_{1}$ and $\mathcal{A}_{2}$ correspond to the elements of the main diagonals of $U_{0}$, $U_{1}$ and $U_{2}$, respectively. When confirming the pure states (FC patterns), a different run may induce a different form of the operator. Therefore, in each stage, we repeat 20 runs and then obtain 20 operators with 20 corresponding eigenvalue spectra. We compare the eigenvalue spectra between different runs, and the corresponding results are shown in Fig. 1.

\begin{figure*}[!t]
	\centering
	\begin{minipage}[t]{0.49\textwidth}
		\centering
		\includegraphics[width=8.6cm]{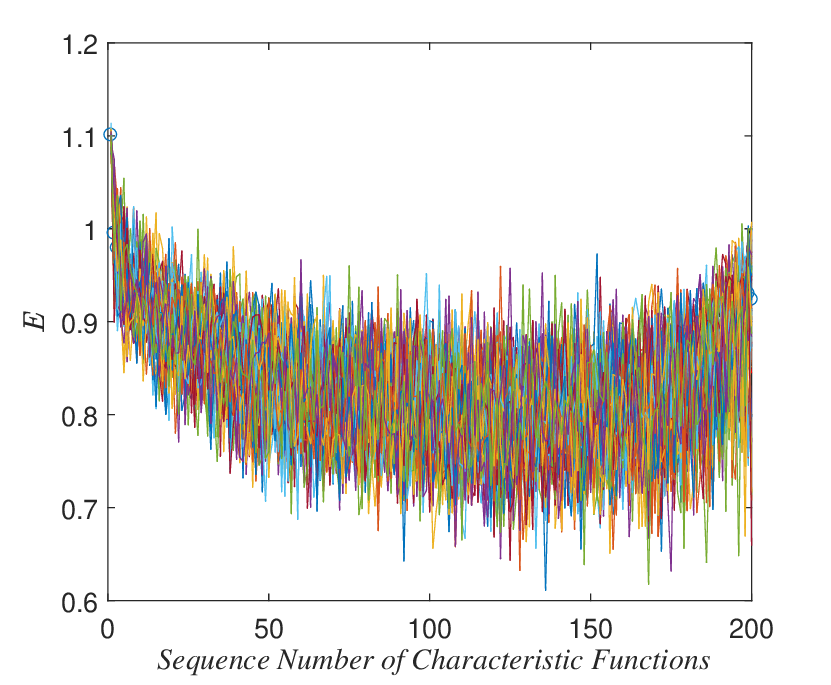}
	\end{minipage}
	\begin{minipage}[t]{0.49\textwidth}
		\centering
		\includegraphics[width=8.6cm]{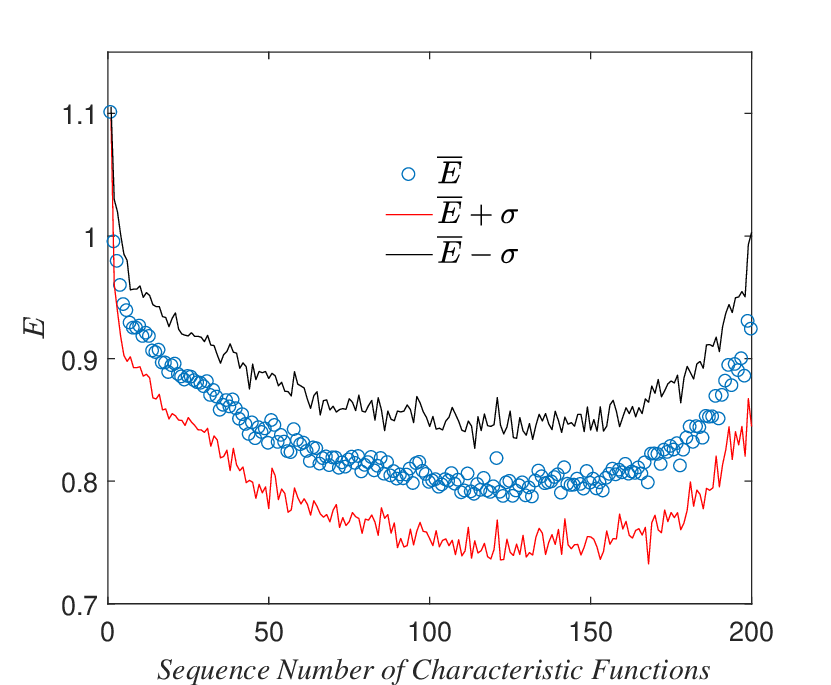}
	\end{minipage}
	\caption{(Left) Eigenvalue spectra of the 60 operators,and (Right) the mean of eigenvalue spectrum $\overline{E}$ by averaging these 60 eigenvalue spectra with the standard deviation $\sigma$.}\label{S_of_A}
\end{figure*}

From Fig. 1, one can see that in general, all the spectra show a consistent trend, and the dispersion $\sigma$ is significantly small. Hence, it can be concluded that the eigenvalue spectra can be seen to remain invariable, regardless of the run or the stage, i.e.,
\begin{equation}
U_{0} = U_{1} = U_{2}.
\end{equation}

\subsubsection{The relation of mechanic operators of $\mathcal{A}_{0}$, $\mathcal{A}_{1}$ and $\mathcal{A}_{2}$}
From part $1$, we obtain
\begin{equation}\nonumber
\mathcal{A}_{0} \sim U_{0}, \mathcal{A}_{1} \sim U_{1}, \mathcal{A}_{2} \sim U_{2},
\end{equation}
and from part $2$, we obtain
\begin{equation}\nonumber
U_{0} = U_{1} = U_{2} 
\end{equation}

Based on the transitivity property and symmetry property of the similarity matrix, we have
\begin{equation}
\mathcal{A}_{0} \sim \mathcal{A}_{1} \sim  \mathcal{A}_{2} 
\end{equation}

Therefore, one can conclude that any two of these three operators are similarity matrices to each other. This means that mechanic operators $\mathcal{A}_{0}$, $\mathcal{A}_{1}$ and $\mathcal{A}_{2}$ can be seen as three different representations of the same linear map under three different bases.

\subsection{Hilbert space of the human brain}

The confirmation of the mechanical operator allows us to build a Hilbert space that is spanned by the operator's eigenvectors. In Hilbert space, any brain state can be presented as a linear combination of the base vectors in the form of a vector in which the elements are expansion coefficients. Since the base vectors are actually derived from the microstates of the brain, they can therefore be seen as the fundamental constituents of the brain state. Then, the studies related to brain state can be refocused on studying fundamental states. For example, the comparison between two brain states in Hilbert space is actually the analysis of how the brain state is presented by the fundamental states. In general, the confirmation of mechanic operators has opened up large possibilites with quantum mechanics to investigate the human brain. In addition to brain state, the state transition, phase transition, etc., could be investigated by means of relevant theories from quantum mechanics. Due to space limitations, further study on Hilbert space based on the current data is not performed. However, the study from this perspective has already been started as another important research project.

\section{Discussions and Conclusions}

In the present work, we proposed a model based on statistical physics. The FC pattern is defined as the microstate of the brain, and the SFC is taken as the macrostate whose propertities can be seen as the statistical average of the microproperties. By applying {\it equilibrium ensemble theory}, the thermodynamic properties of the brain are presented as functions of FC patterns. We applied this model to investigate the biological effect of ketamine on the brain by analyzing the thermodynamic properties. The {\it maximum work principle} based on the decrease in free energy explained the distinct difference between the ketamine and control groups: the maximal work that the brain can do evaluated from the control group is remarkably larger than that evaluated from the ketamine group. The {\it work done} by the brain is suggested to represent the ability of the brain to process information and perform tasks. Hence, it is concluded that the capability of {\it work done} by brain in the control group is remarkably greater than that in the ketamine group.

Taking the microstate of the brain as a mixed quantum state, the mechanical operator of the brain was further investigated. It is found that the mechanical operators confirmed for three stages based on the data from the ketamine group, are actually the different representations of the same linear map under different bases. This result can be expained by the fact that the macrostates of the brain at different stages are basically consistent. The study of the mechanic operator brings a new research perspective for the study of neuroscience: the confirmation of the mechanic operator enables us to build a Hilbert space in which brain states and other properties can be studied by applying quantum mechanics; in addition to the operator that we have obtained in the current study, there exist other operators that describe different dynamic properties of the brain system.

Though there exist some limitations, e.g., reasonabeness of the assumption of brain energy, our model has provided new insight into understanding the human brain. It could also be extensively applied to the diagnosis of some mental diseases, e.g., by monitoring changes in thermodynamic functions. Moreover, since the model is purely dependent on the property of the time series, in addition to the brain system, it is still feasible for other systems when the relevant time series meet the conditions specified in the model.

\end{document}